\documentclass{aastex63}
\usepackage{graphicx} 
\begin{document}

\begin{center}
        \large
       \textbf{NANCY:\\Next-generation All-sky Near-infrared Community surveY}

       \vspace{0.5cm}
       \normalsize
       \begin{flushleft}
        \textbf{Roman Core Community Survey Category:} High Latitude Wide Area Survey
        \vspace{0.1cm}
        
        \normalsize
        \textbf{Scientific Categories:} \textit{Solar system astronomy; stellar physics and stellar types; stellar populations and the interstellar medium; galaxies; the intergalactic medium and the circumgalactic medium, supermassive black holes and active galaxies; large scale structure of the universe}
        \vspace{0.1cm}

        \normalsize
        \textbf{Submitting Authors}: Jesse Han $<$jesse.han@cfa.harvard.edu$>$, Arjun Dey $<$arjun.dey@noirlab.edu$>$
        \vspace{0.1cm}
        
        \textbf{Affiliation}: Harvard-Smithsonian Center for Astrophysics, NOIRLab
        \end{flushleft}
        
        \textbf{Contributing Authors}
   \end{center}
\vspace{-0.8cm}
\author[0000-0002-6800-5778]{Jiwon Jesse Han}
  \affiliation{Center for Astrophysics $|$ Harvard $\&$ Smithsonian, 60 Garden St, Cambridge, MA 02138}
\author[0000-0002-4928-4003]{Arjun Dey}
  \affiliation{NSF's National Optical-Infrared Astronomy Research Laboratory, 950 N Cherry Ave, Tucson, AZ 85719}
\author[0000-0003-0872-7098]{Adrian M.~Price-Whelan}
  \affiliation{Center for Computational Astrophysics, Flatiron Institute, 162 Fifth Ave, New York, NY 10010, USA}
\author[0000-0002-5758-150X]{Joan Najita}
  \affiliation{NSF's National Optical-Infrared Astronomy Research Laboratory, 950 N Cherry Ave, Tucson, AZ 85719}
\author[0000-0002-3569-7421]{Edward F.~Schlafly}
  \affiliation{Space Telescope Science Institute, 3700 San Martin Dr, Baltimore, MD 21218, USA}
\author[0000-0002-6561-9002]{Andrew Saydjari}
  \affiliation{Center for Astrophysics $|$ Harvard $\&$ Smithsonian, 60 Garden St, Cambridge, MA 02138}
\author[0000-0003-2229-011X]{Risa H.~Wechsler}
  \affiliation{Kavli Institute for Particle Astrophysics $\&$ Cosmology, P. O. Box 2450, Stanford University, Stanford, CA 94305, USA }
  \affiliation{Department of Physics, Stanford University, 382 Via Pueblo Mall, Stanford, CA 94305, USA}
  \affiliation{SLAC National Accelerator Laboratory, Menlo Park, CA 94025, USA }
\author[0000-0002-7846-9787]{Ana Bonaca}
  \affiliation{The Observatories of the Carnegie Institution for Science, 813 Santa Barbara Street, Pasadena, CA 91101, USA}
\author[0000-0002-5042-5088]{David J Schlegel}
  \affiliation{Lawrence Berkeley National Lab, 1 Cyclotron Rd, Berkeley CA 94720, USA}
\author{Charlie Conroy}
  \affiliation{Center for Astrophysics $|$ Harvard $\&$ Smithsonian, 60 Garden St, Cambridge, MA 02138}
% ...
\newcommand{\allacks}{
Acknowledgements
}

\section*{Survey Scope}
The Nancy Grace Roman Space Telescope (\textit{Roman}) is capable of delivering an unprecedented \textbf{all-sky, high-spatial resolution, multi-epoch infrared map} to the astronomical community. This opportunity arises in the midst of numerous ground- and space-based surveys that will provide extensive spectroscopy and imaging together covering the entire sky (such as Rubin/LSST, Euclid, UNIONS, SPHEREx, DESI, SDSS-V, GALAH, 4MOST, WEAVE, MOONS, PFS, UVEX, NEO Surveyor, etc.). \textit{Roman} can {\underline{\it uniquely}} provide uniform high-spatial-resolution ($\approx$0.1”) imaging over the entire sky, vastly expanding the science reach and precision of all of these near-term and future surveys. This imaging will not only enhance other surveys,
but also facilitate completely new science. By imaging the full sky over two epochs, \textit{Roman} can measure the proper motions for stars across the entire Milky Way, probing 100 times fainter than \textit{Gaia} out to the very edge of the Galaxy. Here, we propose NANCY: a completely public, all-sky survey that will create a high-value legacy dataset benefiting innumerable ongoing and forthcoming studies of the universe.  NANCY is a pure expression of Roman's potential: it images the entire sky, at high spatial resolution, in a broad infrared bandpass that collects as many photons as possible.  The majority of all ongoing astronomical surveys would benefit from incorporating observations of NANCY into their analyses, whether these surveys focus on nearby stars, the Milky Way, near-field cosmology, or the broader universe.

\textbf{In this white paper, we propose using \textit{Roman} to scan the entire sky to a $5\sigma$ ($10\sigma$) single-exposure depth of 25 AB mag (24.5 mag) in at least two epochs.} There are several survey designs that can reach this goal within the scope of a Core Community Survey (CCS), which we outline in Table \ref{table}. We present three ``fiducial'' designs that we consider equally, and two ``reduced'' designs that are smaller in scope than the fiducial surveys. We note that all of these surveys are significantly shorter than the notional High-Latitude Wide-Area Survey, which is planned to take 24 months (730 days) of on-sky time. We nominally adopt F146 ($\lambda_{\rm cen}=1.464\mu m, \Delta\lambda=1.073\mu m$) as the primary filter due to its high throughput, but also provide how the total survey time would scale up in other filters in Figure \ref{fig:exptime}. In one of the fiducial surveys, we describe an option to add a second filter in F087, F106, or F129. \textbf{In any survey design, completing the first epoch as early in the mission as possible, and repeating the observations near the end of the baseline mission is critical.} This strategy will allow the survey to deliver precise proper motions for faint stars across the entire sky. We describe the fiducial surveys in detail in Section \ref{sec:design}, and the reduced surveys in Section \ref{sec:reduce}.

\begin{figure*}[t]
    \centering
    \includegraphics[width=0.7\textwidth]{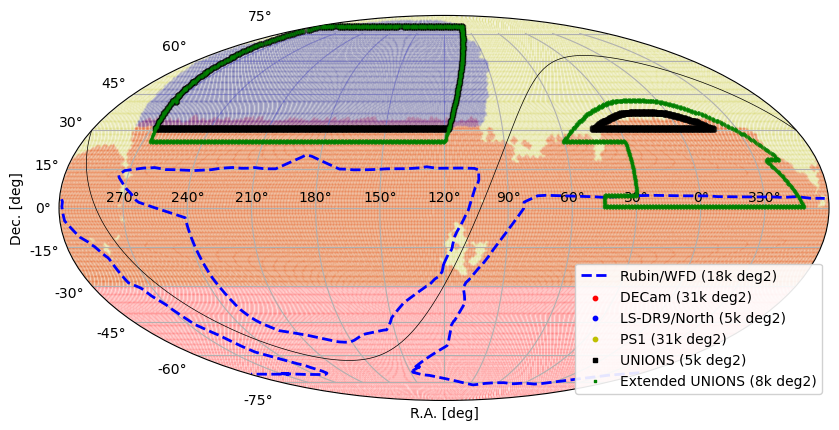}
    \caption{Completed and on-going ground-based imaging surveys cover the entire sky in multiple bands. The figure shows the footprints of the completed PanSTARRS1 \citep{PS1}, the ongoing UNIONS project \citep{CFIS_UNIONS}, existing data from DECam in the NOIRLab archive, and the Legacy Surveys DR9-North survey \citep{LS}. The outline of the Rubin/LSST WFD survey is shown by the blue dashed line. NANCY's $\approx0.1$~arcsec spatial resolution infrared map will cover the \emph{entire sky}, completely overlapping all of these surveys and greatly expanding the value and science of them all.}
    \label{fig:groundbasedcoverage}
\end{figure*}

\section{Science Impact}

Observational cosmology is in its heyday, with a multitude of experiments making precision measurements of the geometry and expansion history of the universe (e.g., eROSITA, SDSS, DES, DESI, Rubin/LSST, Euclid, SPHEREx, CMBS-4). Simultaneously, Galactic astronomy is undergoing a renaissance with the advent of \textit{Gaia}, time-domain surveys such as ATLAS, ZTF and TESS, and the proliferation of highly multiplexed spectroscopic facilities like DESI and 4MOST. In the near-future, Rubin/LSST will be a weak-lensing, cluster-finding, near-field cosmology, and Galactic astronomy engine. Several ongoing and future space missions will result in mapping the sky at X-ray (SRG/eROSITA), UV (UVEX in planning), optical and IR (Euclid in 2023, and SPHEREx in 2025). In the radio, the combination of the VLA All-Sky Survey \citep[VLASS;][]{Lacy20} and ASKAP's EMU survey \citep{norris11} will cover the entire sky in multiple epochs. \textit{All} of these surveys will benefit from 0.1" NIR resolution imaging to identify counterpart sources and their precise morphologies. As the only all-sky, 0.1" resolution infrared survey, NANCY will indeed be \textbf{the} preeminent dataset to fully leverage the investment in other wide-field imaging, spectroscopic and multi-wavelength ground- and space-based missions. For example, the combination of NANCY and Rubin/LSST data can deliver more accurate photo-$z$s and mass measurements of the eROSITA selected galaxy clusters out to redshifts of $z\sim2$, which will constrain the evolution of dark energy. NANCY can also provide SPHEREx \citep{crill20} with a deep prior catalog in the NIR, which will greatly improve the prior-driven SPHEREx spectroscopy and target selection. By the time \textit{Roman} launches, DESI \citep{DESI22} will have measured $\approx$40 million redshifts and 10 million radial velocities; NANCY will yield near-IR morphologies and photometry of every galaxy/QSO with a measured redshift, and proper motions for every star observed by DESI. The legacy value of NANCY will span the whole spectrum in wavelength and time.

At its core, NANCY will produce the highest resolution image of the entire sky in the near-IR (NIR). From this all-sky image, we highlight three Key Products (detailed below):
(1) the highest resolution image of nearly every object in the sky, enabling accurate source classification for ground-based telescopes down to 25 mag in J/H band and accurate weak-lensing shape measurements; 
(2) proper motion measurements to unprecedented depths, enabling kinematic studies of the Galaxy that can pin down the nature of dark matter;  
and (3) a high-spatial-resolution-NIR map of the Milky Way, enabling precision cosmology via improved foreground modeling. Outside of these Key Products, NANCY will broaden the horizon for virtually all aspects of astronomy, including (but not limited to) detecting strong gravitational lenses, establishing a first epoch for transient phenomena, searching for wandering intermediate-mass black holes, discovering low-surface-brightness features such as merging galaxy shells, searching for extragalactic globular clusters and dwarf galaxy companions, and providing robust identifications for SN progenitors in nearby galaxies. We emphasize that NANCY is the \textit{only} opportunity to deliver such a legacy dataset in the foreseeable future.

\subsection{\textbf{Key Product 1}: The highest spatial resolution image of the entire sky}

NANCY will create the highest resolution image of nearly every object in the sky. The most basic use of these data will be the identification of point sources (e.g., stars and quasars) and resolved objects (e.g., galaxies), as well as blended groups of multiple sources. This information is critical throughout astronomy. For instance, ongoing and forthcoming weak-lensing cosmology experiments rely on the robust identification and shape measurements of faint galaxies, but suffer from star–galaxy separation and source confusion issues. Even Rubin/LSST will have star-galaxy confusion at $i\approx24$~AB mag (see, e.g., Fig~4, 5 of Bechtol et al. Roman White paper). The proposed F146 $5\sigma$ depth of 25 AB mag means that NANCY can readily separate point/extended sources brighter than this limit \citep[e.g.,][]{slater20}, which is $>1$~mag fainter than feasible with only the Rubin data. Therefore, NANCY will expand the reaches of the Rubin cosmology program. \textit{Rubin} and Euclid will also create considerable synergies.
In the 15,000 sq. degrees of the Euclid Wide Survey \citep{euclid22}, Euclid will also deliver $\sim0.1"$ resolution imaging in the VIS filter ($0.55-0.9\mu m$, $5\sigma$ at 26.2 mag) that can be combined with NANCY to measure colors of these sources, which greatly enhances the legacy of both surveys. In the NIR, Euclid will observe shallower ($5\sigma$ at 24.5 mag) at lower resolution (0.3” vs. 0.1”) than \textit{Roman}, but still provide an invaluable third epoch to most NANCY sources. Due to its optimal NIR imaging, \textit{Roman} is uniquely suited to solve the problem of source classification for high-redshift sources. Faint source classification also critically impacts near-field science, such as searching for ultra-faint dwarf galaxies tracing the low-mass end of the halo mass function \citep[e.g.,][]{drlica-wagner15}, or studying rare stellar objects in the Galaxy. 
With NANCY, \textit{Roman} presents an opportunity to truly ``resolve'' the source classification problem for all ground-based telescopes.
The \textit{only} competition to NANCY in the forseeable future would be larger NANCY-like programs using \textit{Roman}---and these would benefit tremendously from initial all-sky observations made as part of NANCY.

\subsection{\textbf{Key Product 2}: All-sky astrometry of the lowest mass and most distant stars}

The \textit{Gaia} mission \citep{gaia16} has showcased the impact that precision astrometry can have throughout Galactic astronomy. \textit{Roman} will detect stars much further away and fainter than \textit{Gaia}. As an example, a red clump (RC) star at the Galaxy's virial radius ($\sim250\text{kpc}$) is 22 mag in the F146 filter. NANCY will detect this star with an SNR of 80.  The same star is 23.5 mag in the \textit{Gaia} G band, which is invisible to \textit{Gaia} (limiting magnitude G=21). Beyond the Galaxy, the tip of the red-giant branch will be visible out to $\sim10\text{ Mpc}$ at the $5\sigma$ 25 mag limit of NANCY, covering the entire Local Group including M31, M33, and numerous dwarf galaxies. While it is yet challenging to accurately estimate the proper motion uncertainty from two epochs with a 5-year baseline \citep[the][estimate $25~\mu$as/year for the High-latitude Wide-Area survey]{sanderson19}, we estimate a rough proper motion uncertainty as $\delta_{PM} \sim \text{max} ( \sqrt{2}\times\text{FWHM}/\text{SNR}/\text{5 years}, 0.01\text{pix} / \text{5 years})$, assuming a systematic floor of 0.01pix. From this we infer that NANCY will provide $\sim 200~\mu$as/year precision proper motion measurements for all stars brighter than 21.8 mag in F146, and a precision that scales inversely with SNR down to the faint limit of 25 mag.
In the region of overlap with Euclid, the Euclid imaging data will provide a useful first epoch, increasing the proper motion baseline by a maximum of a factor of $\sim2$. GO observations can also provide more epochs that can reduce the systematic uncertainty floor by $1/\sqrt{N_{\text{epoch}}}$. In combination with millions of radial velocity and chemical abundance measurements from large ground-based spectroscopic surveys (e.g., DESI, SDSS-V, WEAVE, PFS, 4MOST), these measurements will allow us to reveal the formation history of the Milky Way in exquisite detail, and measure density and velocity perturbations in stellar streams imprinted by past encounters with dark matter subhalos.
\vspace{0.2cm}

\subsection{\textbf{Key Product 3}: A High-Spatial-Resolution-NIR Map of the Milky Way}

\textit{Roman}’s high angular resolution and deep NIR sensitivity enable an unprecedented view of the entire Milky Way, including regions with extreme crowding and extinction. Peering through dust to identify globular clusters in the Galactic plane and embedded young stellar objects (YSOs), NANCY will explore star formation in diverse environments from the densest regions of the Galactic center to the spiral arms. By providing the astrometry for a significant fraction of stars in the Galaxy, NANCY will establish the leading catalog for identifying sources of transients, performing forced-photometry on lower-resolution imaging, and targeting spectroscopic surveys, especially those in H-band, such as SDSS-V and MOONS, which are mapping the chemical evolution of the Milky Way \cite[][]{Kollmeier2017, taylor18, almeida23}. Studies of dust cloud morphology at high angular resolution will probe the mechanisms of molecular cloud assembly and evolution as well as the turbulent structure of the ISM across orders of magnitude in spatial scale. The deep F146 photometry will provide “background” sources both at unprecedented distances and for lines of sight with extremely high column densities of dust that, when combined with deep, all-sky, ground-based, multi-color photometry, will extend 3D dust maps to larger distances and into denser cores of molecular clouds \citep[e.g.,][]{green19}. These stellar-reddening based maps of Galactic dust are imperative for precision cosmology to remove Galactic foregrounds in dereddening extragalactic sources at high angular resolution ($\leq1$ arcmin).
\begin{figure*}[t]
    \centering
    \includegraphics[width=0.6\textwidth]{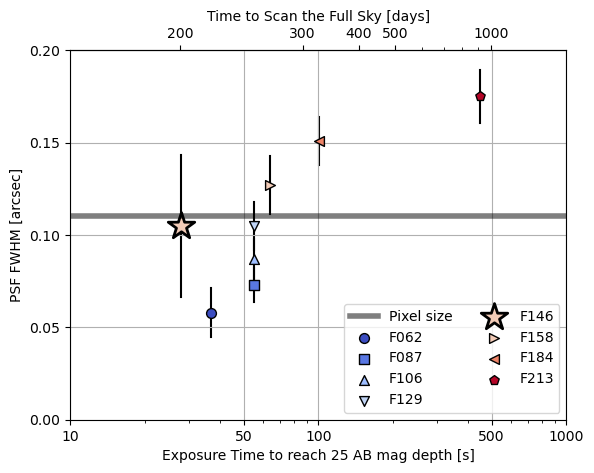}
    \caption{The PSF size in each WFI filter versus the exposure time required to reach 25 AB mag. On the top axis, we also show the time it takes in each filter to scan the whole sky once (assuming 3 dithers, see Section \ref{sec:design}). The range of PSF width resulting from the width of the filter is represented by the vertical bars. The Roman pixel scale of 0.11" is  represented by the thick grey horizontal line.}
    \label{fig:exptime}
\end{figure*}

\section{Fiducial Survey Design} \label{sec:design}

\textit{Roman}'s 0.28 sq.~deg.\ camera can map the entire sky (4$\pi$ sr) in 147,332 contiguous fields. Using the exposure time calculator, we estimate that a total exposure of 30s per field is sufficient to reach the desired $5\sigma$ depth of 25 AB mag in F146, assuming zodiacal light twice greater than minimum. In Figure \ref{fig:exptime} we show the different exposure times required to achieve the same depth in each filter. Based on these basic parameters, we present three possible designs for the fiducial survey. 

In the first design, each exposure is broken up into three 10s dithers (with a 20s slew time between dithers) in order to fill in the chip gaps and improve the PSF sampling. After one field is completed, \textit{Roman} will slew along its short FoV axis (0.4 deg) to the next contiguous field, taking $\approx$50 seconds. Thus, each single-epoch field will take 30s (exposure) + 40s (dither) + 50s (slew) = 120s. This equates to 204 days for one epoch of full-sky, and 408 days for two epochs. We further partition each full-sky scan into two equal (102-day) campaigns: the high-Galactic-latitude ($|b|>30$ deg, 50\% of the full sky) campaign, and the low-latitude ($|b|<30$ deg, 50\% of the full sky) campaign. This partition allows for more flexible scheduling within a given year. It is imperative that the two epochs of a given field are taken at the same time of year; this will minimize the parallax-induced astrometric shifts and allow for accurate proper motion measurements. While this survey design has the benefit that each exposure is well-sampled and void of chip gaps, it is also not an efficient design: the shutter is open only 25\% of the survey time. It may be possible to utilize the slew time to spread out the brightest targets, which can help improve astrometry along the axis perpendicular to the slew. However, the prospect of such “slew exposures” depends greatly on the data volume constraints, and we do not make quantitative estimates here.

In the second design, we do not dither during a single epoch, but instead do three tilings of the sky, with non-overlapping pointing centers for each tiling to fill in the chip gaps. Each single-epoch field takes 30s (exposure) + 50s (slew) = 80s. This strategy allows us to add a \textbf{third} epoch at Y2.5, while still taking 408 days of total survey time, identical to the first design. By placing this intermediate epoch precisely 6 months after 2 years, we can constrain the parallax as well as the proper motion. The astrometric solution for stars that lie in the chip gaps will be more complicated, but even such stars will have at least a 2.5 year baseline for proper motions. In this design, we rely on up-the-ramp sampling for cosmic ray rejection.

A unique opportunity in this three-epoch design is that we can use a different filter for the intermediate epoch. This is the third survey design. For example, replacing the F146 with F062 in the Y2.5 epoch will add just 12 days (420 days total), and F087, F106, or F129 will equally add 44 days to the survey (452 days total). Since F146 is a wide-band filter, adding a narrower filter such as F106 or F129 would serve as a complementary catalog for source classification while also adding color information.

\begin{table}[t]
\begin{center}
\begin{tabular}{lcccccccc}
\hline
Survey &  PM & Total duration & Campaign duration & \multicolumn{5}{c}{Campaigns}\\
& & [days]  &  [days] & Y1 & Y2 & Y2.5 & Y4 & Y5\\
\hline
\hline
Fiducial (Dither) & AL & \textbf{408} & 102 & HL & LL  & - & LL & HL\\
Fiducial (No Dither) & AL & \textbf{408} & 136 & AL & - & AL & - & AL\\
Fiducial (No Dither, Two-band) & AL & \textbf{452} & 136/179 & AL & - & AL [179 days] & - & AL\\
\hline
\hline
3/4 Reduced & HL & \textbf{306} & 102 & HL & LL & - & - & HL\\
Half Reduced & - & \textbf{204} & 102 & HL & LL & - & - & -\\
\hline
\end{tabular}
\end{center}
\caption{Survey Options for implementing NANCY. The top three are the ``fiducial'' designs that we consider equally, and the last two are the ``reduced'' designs that take significantly less time than the fiducial surveys. We note that all of these surveys are significantly shorter than the notional High-Latitude Wide-Area Survey, which is estimated to take 24 months (730 days) of on-sky time. All exposure times assume a single-exposure depth of $5\sigma$ at 25 AB mag. The F146 filter is used in all but the Y2.5 Campaign of the Third Fiducial survey, which uses F087/F102/F126. Column 2 denotes the survey area over which proper motions will be derived. Column 3 is the total duration of the entire survey, and column 4 is the duration for each year-long Campaign. In columns 2, 5-9, AL/HL/LL = All/High/Low Latitudes.}
\label{table}
\end{table}

\section{Design Trade Space} \label{sec:reduce}

The three main parameters of the survey are: survey area (41,253 sq. deg), exposure time (30s), and the F146 filter. As more than half of the survey time is already dedicated to slew/settle, reducing the exposure time would yield a less efficient survey. The choice of the F146 filter is driven by its throughput and sampling. In Figure 1 we plot the exposure time required to reach $5\sigma$ at 25 mag, and the PSF size in arcseconds for each filter. F146 requires the least exposure time. The second fastest filter is F062 ($\lambda_{\rm cen}=0.62\mu m, \Delta\lambda= 0.28\mu m$) at 37s; however, F062 is less well sampled than F146 (0.058” vs. 0.105” PSF, while the pixel scale is 0.11”), which makes the morphological classification of sources at the faint end more difficult. Thus, F146 is the ideal filter for a fast survey. Since reducing the exposure time is not viable, and using a different filter only increases the total survey time, the survey area offers the only meaningful trade space to reduce the total survey time. We thus present two ``reduced'' surveys that are smaller in survey area.

The first reduced survey is the “3/4” survey, which scans the high-latitude (HL) sky twice and the low-latitude (LL) sky only once (HL in Y1, LL spread over Y2–4, HL in Y5). This will still enable astrometry in the halo and extinction mapping in the Galactic plane. The second reduced survey is the “half” survey, which scans the full sky once in Y1. This design will rely on GO programs to obtain astrometry for targeted fields, but will still deliver an invaluable all-sky source catalog. \textbf{In all cases, it is still critical that, at the minimum,  an all-sky, high-Galactic-latitude campaign occurs in Y1 in order to enable precise astrometric measurements later during the mission.}

While in this white paper we have focused on delivering the fastest possible full-sky survey, there are several reasons to consider a longer, more full-fledged version of NANCY. First, a key challenge to designing any \textit{Roman} survey is the considerable overhead (20s per chip-gap dither, 50s per short-FoV axis slew). A direct way to make the survey more efficient would be to increase the exposure times, thereby reducing the fraction of time spent on slew/settle. For example, if we entirely replaced F146 with F087, F106, or F129, the required exposure time to reach the $5\sigma$ depth at 25 mag is 55s, and the fiducial survey with dithering would take 491 days. Using these narrower filters would also remedy a major disadvantage of the F146 filter that it is difficult to accurately calculate the PSF for sources of unknown color due to the exceptional wavelength range. One could even consider a multi-epoch, multi-band all-sky survey that would provide colors for every object in the survey; such a survey would likely take $\sim900$ days of survey time. To improve proper motion uncertainties, adding additional epochs is more effective than taking deeper exposures, as the astrometric measurements are limited by systematics for most sources of interest in the Galaxy. The same strategy would benefit the Time Domain CCS as well: increasing the number of epochs directly increases the discovery space in time. 

\section{Summary} \label{sec:summary}
\textit{Roman} is the \textbf{only} instrument capable of carrying out a high-resolution, all-sky NIR survey like NANCY for the foreseeable future. The legacy value of this survey spans the whole spectrum of astronomy, and ties together the innumerable small- and large-scale investments going into ground and space-based telescopes. Furthermore, NANCY will only improve the science yield of other \textit{Roman} programs, including its core cosmology program (e.g., by creating synergies with Rubin, Euclid and DESI) and its time domain program (by establishing an all-sky single epoch baseline). \textbf{We thus urge the \textit{Roman} CCS committee to broaden the scope of the High-Latitude Wide-Area Survey to include, or expand upon, NANCY.}

\bibliography{refs}{}

\begin{thebibliography}{}
\expandafter\ifx\csname natexlab\endcsname\relax\def\natexlab#1{#1}\fi
\providecommand{\url}[1]{\href{#1}{#1}}
\providecommand{\dodoi}[1]{doi:~\href{http://doi.org/#1}{\nolinkurl{#1}}}
\providecommand{\doeprint}[1]{\href{http://ascl.net/#1}{\nolinkurl{http://ascl.net/#1}}}
\providecommand{\doarXiv}[1]{\href{https://arxiv.org/abs/#1}{\nolinkurl{https://arxiv.org/abs/#1}}}

\bibitem[{{Almeida} {et~al.}(2023){Almeida}, {Anderson},
  {Argudo-Fern{\'a}ndez}, {Badenes}, {Barger}, {Barrera-Ballesteros}, {Bender},
  {Benitez}, {Besser}, {Bizyaev}, {Blanton}, {Bochanski}, {Bovy}, {Brandt},
  {Brownstein}, {Buchner}, {Bulbul}, {Burchett}, {Cano D{\'\i}az}, {Carlberg},
  {Casey}, {Chandra}, {Cherinka}, {Chiappini}, {Coker}, {Comparat}, {Conroy},
  {Contardo}, {Cortes}, {Covey}, {Crane}, {Cunha}, {Dabbieri}, {Davidson},
  {Davis}, {De Lee}, {M{\'e}ndez Delgado}, {Demasi}, {Di Mille}, {Donor},
  {Dow}, {Dwelly}, {Eracleous}, {Eriksen}, {Fan}, {Farr}, {Frederick}, {Fries},
  {Frinchaboy}, {Gaensicke}, {Ge}, {Gonz{\'a}lez {\'A}vila}, {Grabowski},
  {Grier}, {Guiglion}, {Gupta}, {Hall}, {Hawkins}, {Hayes}, {Hermes},
  {Hern{\'a}ndez-Garc{\'\i}a}, {Hogg}, {Holtzman}, {Ibarra-Medel}, {Ji},
  {Jofre}, {Johnson}, {Jones}, {Kinemuchi}, {Kluge}, {Koekemoer}, {Kollmeier},
  {Kounkel}, {Krishnarao}, {Krumpe}, {Lacerna}, {Jakson Assuncao Lago},
  {Laporte}, {Liu}, {Liu}, {Liu}, {Lopes}, {Macktoobian}, {Malanushenko},
  {Maoz}, {Masseron}, {Masters}, {Matijevic}, {McBride}, {Medan}, {Merloni},
  {Morrison}, {Myers}, {M{\'e}sz{\'a}ros}, {Negrete}, {Nidever}, {Nitschelm},
  {Oravetz}, {Oravetz}, {Pan}, {Peng}, {Pinsonneault}, {Pogge}, {Qiu},
  {Queiroz}, {Ramirez}, {Rix}, {Fern{\'a}ndez Rosso}, {Runnoe}, {Salvato},
  {Sanchez}, {Santana}, {Saydjari}, {Sayres}, {Schlaufman}, {Schneider},
  {Schwope}, {Serna}, {Shen}, {Sobeck}, {Song}, {Souto}, {Spoo}, {Stassun},
  {Steinmetz}, {Straumit}, {Stringfellow}, {S{\'a}nchez-Gallego},
  {Taghizadeh-Popp}, {Tayar}, {Thakar}, {Tissera}, {Tkachenko}, {Hernandez
  Toledo}, {Trakhtenbrot}, {Fernandez Trincado}, {Troup}, {Trump}, {Tuttle},
  {Ulloa}, {Vazquez-Mata}, {Alfaro}, {Villanova}, {Wachter}, {Weijmans},
  {Wheeler}, {Wilson}, {Wojno}, {Wolf}, {Xue}, {Ybarra}, {Zari}, \&
  {Zasowski}}]{almeida23}
{Almeida}, A., {Anderson}, S.~F., {Argudo-Fern{\'a}ndez}, M., {et~al.} 2023,
  arXiv e-prints, arXiv:2301.07688, \dodoi{10.48550/arXiv.2301.07688}

\bibitem[{{Chambers} {et~al.}(2016){Chambers}, {Magnier}, {Metcalfe},
  {Flewelling}, {Huber}, {Waters}, {Denneau}, {Draper}, {Farrow}, {Finkbeiner},
  {Holmberg}, {Koppenhoefer}, {Price}, {Rest}, {Saglia}, {Schlafly}, {Smartt},
  {Sweeney}, {Wainscoat}, {Burgett}, {Chastel}, {Grav}, {Heasley}, {Hodapp},
  {Jedicke}, {Kaiser}, {Kudritzki}, {Luppino}, {Lupton}, {Monet}, {Morgan},
  {Onaka}, {Shiao}, {Stubbs}, {Tonry}, {White}, {Ba{\~n}ados}, {Bell},
  {Bender}, {Bernard}, {Boegner}, {Boffi}, {Botticella}, {Calamida},
  {Casertano}, {Chen}, {Chen}, {Cole}, {Deacon}, {Frenk}, {Fitzsimmons},
  {Gezari}, {Gibbs}, {Goessl}, {Goggia}, {Gourgue}, {Goldman}, {Grant},
  {Grebel}, {Hambly}, {Hasinger}, {Heavens}, {Heckman}, {Henderson}, {Henning},
  {Holman}, {Hopp}, {Ip}, {Isani}, {Jackson}, {Keyes}, {Koekemoer}, {Kotak},
  {Le}, {Liska}, {Long}, {Lucey}, {Liu}, {Martin}, {Masci}, {McLean}, {Mindel},
  {Misra}, {Morganson}, {Murphy}, {Obaika}, {Narayan}, {Nieto-Santisteban},
  {Norberg}, {Peacock}, {Pier}, {Postman}, {Primak}, {Rae}, {Rai}, {Riess},
  {Riffeser}, {Rix}, {R{\"o}ser}, {Russel}, {Rutz}, {Schilbach}, {Schultz},
  {Scolnic}, {Strolger}, {Szalay}, {Seitz}, {Small}, {Smith}, {Soderblom},
  {Taylor}, {Thomson}, {Taylor}, {Thakar}, {Thiel}, {Thilker}, {Unger},
  {Urata}, {Valenti}, {Wagner}, {Walder}, {Walter}, {Watters}, {Werner},
  {Wood-Vasey}, \& {Wyse}}]{PS1}
{Chambers}, K.~C., {Magnier}, E.~A., {Metcalfe}, N., {et~al.} 2016, arXiv
  e-prints, arXiv:1612.05560, \dodoi{10.48550/arXiv.1612.05560}

\bibitem[{{Crill} {et~al.}(2020){Crill}, {Werner}, {Akeson}, {Ashby}, {Bleem},
  {Bock}, {Bryan}, {Burnham}, {Byunh}, {Chang}, {Chiang}, {Cook}, {Cooray},
  {Davis}, {Dor{\'e}}, {Dowell}, {Dubois-Felsmann}, {Eifler}, {Faisst},
  {Habib}, {Heinrich}, {Heitmann}, {Heaton}, {Hirata}, {Hristov}, {Hui},
  {Jeong}, {Kang}, {Kecman}, {Kirkpatrick}, {Korngut}, {Krause}, {Lee},
  {Lisse}, {Masters}, {Mauskopf}, {Melnick}, {Miyasaka}, {Nayyeri}, {Nguyen},
  {{\"O}berg}, {Padin}, {Paladini}, {Pourrahmani}, {Pyo}, {Smith}, {Song},
  {Symons}, {Teplitz}, {Tolls}, {Unwin}, {Windhorst}, {Yang}, \&
  {Zemcov}}]{crill20}
{Crill}, B.~P., {Werner}, M., {Akeson}, R., {et~al.} 2020, in Society of
  Photo-Optical Instrumentation Engineers (SPIE) Conference Series, Vol. 11443,
  Space Telescopes and Instrumentation 2020: Optical, Infrared, and Millimeter
  Wave, ed. M.~{Lystrup} \& M.~D. {Perrin}, 114430I, \dodoi{10.1117/12.2567224}

\bibitem[{{DESI Collaboration} {et~al.}(2022){DESI Collaboration}, {Abareshi},
  {Aguilar}, {Ahlen}, {Alam}, {Alexander}, {Alfarsy}, {Allen}, {Allende
  Prieto}, {Alves}, {Ameel}, {Armengaud}, {Asorey}, {Aviles}, {Bailey},
  {Balaguera-Antol{\'\i}nez}, {Ballester}, {Baltay}, {Bault}, {Beltran},
  {Benavides}, {BenZvi}, {Berti}, {Besuner}, {Beutler}, {Bianchi}, {Blake},
  {Blanc}, {Blum}, {Bolton}, {Bose}, {Bramall}, {Brieden}, {Brodzeller},
  {Brooks}, {Brownewell}, {Buckley-Geer}, {Cahn}, {Cai}, {Canning}, {Capasso},
  {Carnero Rosell}, {Carton}, {Casas}, {Castander}, {Cervantes-Cota},
  {Chabanier}, {Chaussidon}, {Chuang}, {Circosta}, {Cole}, {Cooper}, {da
  Costa}, {Cousinou}, {Cuceu}, {Davis}, {Dawson}, {de la Cruz-Noriega}, {de la
  Macorra}, {de Mattia}, {Della Costa}, {Demmer}, {Derwent}, {Dey}, {Dey},
  {Dhungana}, {Ding}, {Dobson}, {Doel}, {Donald-McCann}, {Donaldson},
  {Douglass}, {Duan}, {Dunlop}, {Edelstein}, {Eftekharzadeh}, {Eisenstein},
  {Enriquez-Vargas}, {Escoffier}, {Evatt}, {Fagrelius}, {Fan}, {Fanning},
  {Fawcett}, {Ferraro}, {Ereza}, {Flaugher}, {Font-Ribera}, {Forero-Romero},
  {Frenk}, {Fromenteau}, {G{\"a}nsicke}, {Garcia-Quintero}, {Garrison},
  {Gazta{\~n}aga}, {Gerardi}, {Gil-Mar{\'\i}n}, {Gontcho a Gontcho},
  {Gonzalez-Morales}, {Gonzalez-de-Rivera}, {Gonzalez-Perez}, {Gordon},
  {Graur}, {Green}, {Grove}, {Gruen}, {Gutierrez}, {Guy}, {Hahn}, {Harris},
  {Herrera}, {Herrera-Alcantar}, {Honscheid}, {Howlett}, {Huterer},
  {Ir{\v{s}}i{\v{c}}}, {Ishak}, {Jelinsky}, {Jiang}, {Jimenez}, {Jing},
  {Joyce}, {Jullo}, {Juneau}, {Kara{\c{c}}ayl{\i}}, {Karamanis}, {Karcher},
  {Karim}, {Kehoe}, {Kent}, {Kirkby}, {Kisner}, {Kitaura}, {Koposov},
  {Kov{\'a}cs}, {Kremin}, {Krolewski}, {L'Huillier}, {Lahav}, {Lambert},
  {Lamman}, {Lan}, {Landriau}, {Lane}, {Lang}, {Lange}, {Lasker}, {Le Guillou},
  {Leauthaud}, {Le Van Suu}, {Levi}, {Li}, {Magneville}, {Manera}, {Manser},
  {Marshall}, {Martini}, {McCollam}, {McDonald}, {Meisner},
  {Mena-Fern{\'a}ndez}, {Meneses-Rizo}, {Mezcua}, {Miller}, {Miquel},
  {Montero-Camacho}, {Moon}, {Moustakas}, {Mueller}, {Mu{\~n}oz-Guti{\'e}rrez},
  {Myers}, {Nadathur}, {Najita}, {Napolitano}, {Neilsen}, {Newman}, {Nie},
  {Ning}, {Niz}, {Norberg}, {Noriega}, {O'Brien}, {Obuljen},
  {Palanque-Delabrouille}, {Palmese}, {Zhiwei}, {Pappalardo}, {PENG},
  {Percival}, {Perruchot}, {Pogge}, {Poppett}, {Porredon}, {Prada},
  {Prochaska}, {Pucha}, {P{\'e}rez-Fern{\'a}ndez}, {P{\'e}rez-R{\`a}fols},
  {Rabinowitz}, {Raichoor}, {Ramirez-Solano}, {Ram{\'\i}rez-P{\'e}rez},
  {Ravoux}, {Reil}, {Rezaie}, {Rocher}, {Rockosi}, {Roe}, {Roodman}, {Ross},
  {Rossi}, {Ruggeri}, {Ruhlmann-Kleider}, {Sabiu}, {Safonova}, {Said},
  {Saintonge}, {Salas Catonga}, {Samushia}, {Sanchez}, {Saulder}, {Schaan},
  {Schlafly}, {Schlegel}, {Schmoll}, {Scholte}, {Schubnell}, {Secroun}, {Seo},
  {Serrano}, {Sharples}, {Sholl}, {Silber}, {Silva}, {Sirk}, {Siudek}, {Smith},
  {Sprayberry}, {Staten}, {Stupak}, {Tan}, {Tarl{\'e}}, {Tie}, {Tojeiro},
  {Ure{\~n}a-L{\'o}pez}, {Valdes}, {Valenzuela}, {Valluri},
  {Vargas-Maga{\~n}a}, {Verde}, {Walther}, {Wang}, {Wang}, {Weaver},
  {Weaverdyck}, {Wechsler}, {Wilson}, {Yang}, {Yu}, {Yuan}, {Y{\`e}che},
  {Zhang}, {Zhang}, {Zhao}, {Zhou}, {Zhou}, {Zou}, {Zou}, {Zou}, {Zu}, \& {DESI
  Collaboration}}]{DESI22}
{DESI Collaboration}, {Abareshi}, B., {Aguilar}, J., {et~al.} 2022, \aj, 164,
  207, \dodoi{10.3847/1538-3881/ac882b}

\bibitem[{{Dey} {et~al.}(2019){Dey}, {Schlegel}, {Lang}, {Blum}, {Burleigh},
  {Fan}, {Findlay}, {Finkbeiner}, {Herrera}, {Juneau}, {Landriau}, {Levi},
  {McGreer}, {Meisner}, {Myers}, {Moustakas}, {Nugent}, {Patej}, {Schlafly},
  {Walker}, {Valdes}, {Weaver}, {Y{\`e}che}, {Zou}, {Zhou}, {Abareshi},
  {Abbott}, {Abolfathi}, {Aguilera}, {Alam}, {Allen}, {Alvarez}, {Annis},
  {Ansarinejad}, {Aubert}, {Beechert}, {Bell}, {BenZvi}, {Beutler}, {Bielby},
  {Bolton}, {Brice{\~n}o}, {Buckley-Geer}, {Butler}, {Calamida}, {Carlberg},
  {Carter}, {Casas}, {Castander}, {Choi}, {Comparat}, {Cukanovaite}, {Delubac},
  {DeVries}, {Dey}, {Dhungana}, {Dickinson}, {Ding}, {Donaldson}, {Duan},
  {Duckworth}, {Eftekharzadeh}, {Eisenstein}, {Etourneau}, {Fagrelius},
  {Farihi}, {Fitzpatrick}, {Font-Ribera}, {Fulmer}, {G{\"a}nsicke},
  {Gaztanaga}, {George}, {Gerdes}, {Gontcho}, {Gorgoni}, {Green}, {Guy},
  {Harmer}, {Hernandez}, {Honscheid}, {Huang}, {James}, {Jannuzi}, {Jiang},
  {Joyce}, {Karcher}, {Karkar}, {Kehoe}, {Kneib}, {Kueter-Young}, {Lan},
  {Lauer}, {Le Guillou}, {Le Van Suu}, {Lee}, {Lesser}, {Perreault Levasseur},
  {Li}, {Mann}, {Marshall}, {Mart{\'\i}nez-V{\'a}zquez}, {Martini}, {du Mas des
  Bourboux}, {McManus}, {Meier}, {M{\'e}nard}, {Metcalfe},
  {Mu{\~n}oz-Guti{\'e}rrez}, {Najita}, {Napier}, {Narayan}, {Newman}, {Nie},
  {Nord}, {Norman}, {Olsen}, {Paat}, {Palanque-Delabrouille}, {Peng},
  {Poppett}, {Poremba}, {Prakash}, {Rabinowitz}, {Raichoor}, {Rezaie},
  {Robertson}, {Roe}, {Ross}, {Ross}, {Rudnick}, {Safonova}, {Saha},
  {S{\'a}nchez}, {Savary}, {Schweiker}, {Scott}, {Seo}, {Shan}, {Silva},
  {Slepian}, {Soto}, {Sprayberry}, {Staten}, {Stillman}, {Stupak}, {Summers},
  {Sien Tie}, {Tirado}, {Vargas-Maga{\~n}a}, {Vivas}, {Wechsler}, {Williams},
  {Yang}, {Yang}, {Yapici}, {Zaritsky}, {Zenteno}, {Zhang}, {Zhang}, {Zhou}, \&
  {Zhou}}]{LS}
{Dey}, A., {Schlegel}, D.~J., {Lang}, D., {et~al.} 2019, \aj, 157, 168,
  \dodoi{10.3847/1538-3881/ab089d}

\bibitem[{{Drlica-Wagner} {et~al.}(2015){Drlica-Wagner}, {Bechtol}, {Rykoff},
  {Luque}, {Queiroz}, {Mao}, {Wechsler}, {Simon}, {Santiago}, {Yanny},
  {Balbinot}, {Dodelson}, {Fausti Neto}, {James}, {Li}, {Maia}, {Marshall},
  {Pieres}, {Stringer}, {Walker}, {Abbott}, {Abdalla}, {Allam},
  {Benoit-L{\'e}vy}, {Bernstein}, {Bertin}, {Brooks}, {Buckley-Geer}, {Burke},
  {Carnero Rosell}, {Carrasco Kind}, {Carretero}, {Crocce}, {da Costa},
  {Desai}, {Diehl}, {Dietrich}, {Doel}, {Eifler}, {Evrard}, {Finley},
  {Flaugher}, {Fosalba}, {Frieman}, {Gaztanaga}, {Gerdes}, {Gruen}, {Gruendl},
  {Gutierrez}, {Honscheid}, {Kuehn}, {Kuropatkin}, {Lahav}, {Martini},
  {Miquel}, {Nord}, {Ogando}, {Plazas}, {Reil}, {Roodman}, {Sako}, {Sanchez},
  {Scarpine}, {Schubnell}, {Sevilla-Noarbe}, {Smith}, {Soares-Santos},
  {Sobreira}, {Suchyta}, {Swanson}, {Tarle}, {Tucker}, {Vikram}, {Wester},
  {Zhang}, {Zuntz}, \& {DES Collaboration}}]{drlica-wagner15}
{Drlica-Wagner}, A., {Bechtol}, K., {Rykoff}, E.~S., {et~al.} 2015, \apj, 813,
  109, \dodoi{10.1088/0004-637X/813/2/109}

\bibitem[{{Euclid Collaboration} {et~al.}(2022){Euclid Collaboration},
  {Scaramella}, {Amiaux}, {Mellier}, {Burigana}, {Carvalho}, {Cuillandre}, {Da
  Silva}, {Derosa}, {Dinis}, {Maiorano}, {Maris}, {Tereno}, {Laureijs},
  {Boenke}, {Buenadicha}, {Dupac}, {Gaspar Venancio}, {G{\'o}mez-{\'A}lvarez},
  {Hoar}, {Lorenzo Alvarez}, {Racca}, {Saavedra-Criado}, {Schwartz}, {Vavrek},
  {Schirmer}, {Aussel}, {Azzollini}, {Cardone}, {Cropper}, {Ealet}, {Garilli},
  {Gillard}, {Granett}, {Guzzo}, {Hoekstra}, {Jahnke}, {Kitching}, {Maciaszek},
  {Meneghetti}, {Miller}, {Nakajima}, {Niemi}, {Pasian}, {Percival},
  {Pottinger}, {Sauvage}, {Scodeggio}, {Wachter}, {Zacchei}, {Aghanim},
  {Amara}, {Auphan}, {Auricchio}, {Awan}, {Balestra}, {Bender}, {Bodendorf},
  {Bonino}, {Branchini}, {Brau-Nogue}, {Brescia}, {Candini}, {Capobianco},
  {Carbone}, {Carlberg}, {Carretero}, {Casas}, {Castander}, {Castellano},
  {Cavuoti}, {Cimatti}, {Cledassou}, {Congedo}, {Conselice}, {Conversi},
  {Copin}, {Corcione}, {Costille}, {Courbin}, {Degaudenzi}, {Douspis},
  {Dubath}, {Duncan}, {Dusini}, {Farrens}, {Ferriol}, {Fosalba}, {Fourmanoit},
  {Frailis}, {Franceschi}, {Franzetti}, {Fumana}, {Gillis}, {Giocoli},
  {Grazian}, {Grupp}, {Haugan}, {Holmes}, {Hormuth}, {Hudelot}, {Kermiche},
  {Kiessling}, {Kilbinger}, {Kohley}, {Kubik}, {K{\"u}mmel}, {Kunz},
  {Kurki-Suonio}, {Lahav}, {Ligori}, {Lilje}, {Lloro}, {Mansutti}, {Marggraf},
  {Markovic}, {Marulli}, {Massey}, {Maurogordato}, {Melchior}, {Merlin},
  {Meylan}, {Mohr}, {Moresco}, {Morin}, {Moscardini}, {Munari}, {Nichol},
  {Padilla}, {Paltani}, {Peacock}, {Pedersen}, {Pettorino}, {Pires}, {Poncet},
  {Popa}, {Pozzetti}, {Raison}, {Rebolo}, {Rhodes}, {Rix}, {Roncarelli},
  {Rossetti}, {Saglia}, {Schneider}, {Schrabback}, {Secroun}, {Seidel},
  {Serrano}, {Sirignano}, {Sirri}, {Skottfelt}, {Stanco}, {Starck},
  {Tallada-Cresp{\'\i}}, {Tavagnacco}, {Taylor}, {Teplitz}, {Toledo-Moreo},
  {Torradeflot}, {Trifoglio}, {Valentijn}, {Valenziano}, {Verdoes Kleijn},
  {Wang}, {Welikala}, {Weller}, {Wetzstein}, {Zamorani}, {Zoubian}, {Andreon},
  {Baldi}, {Bardelli}, {Boucaud}, {Camera}, {Di Ferdinando}, {Fabbian},
  {Farinelli}, {Galeotta}, {Graci{\'a}-Carpio}, {Maino}, {Medinaceli}, {Mei},
  {Neissner}, {Polenta}, {Renzi}, {Romelli}, {Rosset}, {Sureau}, {Tenti},
  {Vassallo}, {Zucca}, {Baccigalupi}, {Balaguera-Antol{\'\i}nez}, {Battaglia},
  {Biviano}, {Borgani}, {Bozzo}, {Cabanac}, {Cappi}, {Casas}, {Castignani},
  {Colodro-Conde}, {Coupon}, {Courtois}, {Cuby}, {de la Torre}, {Desai},
  {Dole}, {Fabricius}, {Farina}, {Ferreira}, {Finelli}, {Flose-Reimberg},
  {Fotopoulou}, {Ganga}, {Gozaliasl}, {Hook}, {Keihanen}, {Kirkpatrick},
  {Liebing}, {Lindholm}, {Mainetti}, {Martinelli}, {Martinet}, {Maturi},
  {McCracken}, {Metcalf}, {Morgante}, {Nightingale}, {Nucita}, {Patrizii},
  {Potter}, {Riccio}, {S{\'a}nchez}, {Sapone}, {Schewtschenko}, {Schultheis},
  {Scottez}, {Teyssier}, {Tutusaus}, {Valiviita}, {Viel}, {Vriend}, \&
  {Whittaker}}]{euclid22}
{Euclid Collaboration}, {Scaramella}, R., {Amiaux}, J., {et~al.} 2022, \aap,
  662, A112, \dodoi{10.1051/0004-6361/202141938}

\bibitem[{{Gaia Collaboration} {et~al.}(2016){Gaia Collaboration}, {Prusti},
  {de Bruijne}, {Brown}, {Vallenari}, {Babusiaux}, {Bailer-Jones}, {Bastian},
  {Biermann}, {Evans}, {Eyer}, {Jansen}, {Jordi}, {Klioner}, {Lammers},
  {Lindegren}, {Luri}, {Mignard}, {Milligan}, {Panem}, {Poinsignon},
  {Pourbaix}, {Randich}, {Sarri}, {Sartoretti}, {Siddiqui}, {Soubiran},
  {Valette}, {van Leeuwen}, {Walton}, {Aerts}, {Arenou}, {Cropper}, {Drimmel},
  {H{\o}g}, {Katz}, {Lattanzi}, {O'Mullane}, {Grebel}, {Holland}, {Huc},
  {Passot}, {Bramante}, {Cacciari}, {Casta{\~n}eda}, {Chaoul}, {Cheek}, {De
  Angeli}, {Fabricius}, {Guerra}, {Hern{\'a}ndez}, {Jean-Antoine-Piccolo},
  {Masana}, {Messineo}, {Mowlavi}, {Nienartowicz}, {Ord{\'o}{\~n}ez-Blanco},
  {Panuzzo}, {Portell}, {Richards}, {Riello}, {Seabroke}, {Tanga},
  {Th{\'e}venin}, {Torra}, {Els}, {Gracia-Abril}, {Comoretto},
  {Garcia-Reinaldos}, {Lock}, {Mercier}, {Altmann}, {Andrae}, {Astraatmadja},
  {Bellas-Velidis}, {Benson}, {Berthier}, {Blomme}, {Busso}, {Carry},
  {Cellino}, {Clementini}, {Cowell}, {Creevey}, {Cuypers}, {Davidson}, {De
  Ridder}, {de Torres}, {Delchambre}, {Dell'Oro}, {Ducourant}, {Fr{\'e}mat},
  {Garc{\'\i}a-Torres}, {Gosset}, {Halbwachs}, {Hambly}, {Harrison}, {Hauser},
  {Hestroffer}, {Hodgkin}, {Huckle}, {Hutton}, {Jasniewicz}, {Jordan},
  {Kontizas}, {Korn}, {Lanzafame}, {Manteiga}, {Moitinho}, {Muinonen},
  {Osinde}, {Pancino}, {Pauwels}, {Petit}, {Recio-Blanco}, {Robin}, {Sarro},
  {Siopis}, {Smith}, {Smith}, {Sozzetti}, {Thuillot}, {van Reeven}, {Viala},
  {Abbas}, {Abreu Aramburu}, {Accart}, {Aguado}, {Allan}, {Allasia},
  {Altavilla}, {{\'A}lvarez}, {Alves}, {Anderson}, {Andrei}, {Anglada Varela},
  {Antiche}, {Antoja}, {Ant{\'o}n}, {Arcay}, {Atzei}, {Ayache}, {Bach},
  {Baker}, {Balaguer-N{\'u}{\~n}ez}, {Barache}, {Barata}, {Barbier}, {Barblan},
  {Baroni}, {Barrado y Navascu{\'e}s}, {Barros}, {Barstow}, {Becciani},
  {Bellazzini}, {Bellei}, {Bello Garc{\'\i}a}, {Belokurov}, {Bendjoya},
  {Berihuete}, {Bianchi}, {Bienaym{\'e}}, {Billebaud}, {Blagorodnova},
  {Blanco-Cuaresma}, {Boch}, {Bombrun}, {Borrachero}, {Bouquillon}, {Bourda},
  {Bouy}, {Bragaglia}, {Breddels}, {Brouillet}, {Br{\"u}semeister},
  {Bucciarelli}, {Budnik}, {Burgess}, {Burgon}, {Burlacu}, {Busonero}, {Buzzi},
  {Caffau}, {Cambras}, {Campbell}, {Cancelliere}, {Cantat-Gaudin}, {Carlucci},
  {Carrasco}, {Castellani}, {Charlot}, {Charnas}, {Charvet}, {Chassat},
  {Chiavassa}, {Clotet}, {Cocozza}, {Collins}, {Collins}, {Costigan}, {Crifo},
  {Cross}, {Crosta}, {Crowley}, {Dafonte}, {Damerdji}, {Dapergolas}, {David},
  {David}, {De Cat}, {de Felice}, {de Laverny}, {De Luise}, {De March}, {de
  Martino}, {de Souza}, {Debosscher}, {del Pozo}, {Delbo}, {Delgado},
  {Delgado}, {di Marco}, {Di Matteo}, {Diakite}, {Distefano}, {Dolding}, {Dos
  Anjos}, {Drazinos}, {Dur{\'a}n}, {Dzigan}, {Ecale}, {Edvardsson}, {Enke},
  {Erdmann}, {Escolar}, {Espina}, {Evans}, {Eynard Bontemps}, {Fabre},
  {Fabrizio}, {Faigler}, {Falc{\~a}o}, {Farr{\`a}s Casas}, {Faye}, {Federici},
  {Fedorets}, {Fern{\'a}ndez-Hern{\'a}ndez}, {Fernique}, {Fienga}, {Figueras},
  {Filippi}, {Findeisen}, {Fonti}, {Fouesneau}, {Fraile}, {Fraser}, {Fuchs},
  {Furnell}, {Gai}, {Galleti}, {Galluccio}, {Garabato}, {Garc{\'\i}a-Sedano},
  {Gar{\'e}}, {Garofalo}, {Garralda}, {Gavras}, {Gerssen}, {Geyer}, {Gilmore},
  {Girona}, {Giuffrida}, {Gomes}, {Gonz{\'a}lez-Marcos},
  {Gonz{\'a}lez-N{\'u}{\~n}ez}, {Gonz{\'a}lez-Vidal}, {Granvik}, {Guerrier},
  {Guillout}, {Guiraud}, {G{\'u}rpide}, {Guti{\'e}rrez-S{\'a}nchez}, {Guy},
  {Haigron}, {Hatzidimitriou}, {Haywood}, {Heiter}, {Helmi}, {Hobbs},
  {Hofmann}, {Holl}, {Holland}, {Hunt}, {Hypki}, {Icardi}, {Irwin}, {Jevardat
  de Fombelle}, {Jofr{\'e}}, {Jonker}, {Jorissen}, {Julbe}, {Karampelas},
  {Kochoska}, {Kohley}, {Kolenberg}, {Kontizas}, {Koposov}, {Kordopatis},
  {Koubsky}, {Kowalczyk}, {Krone-Martins}, {Kudryashova}, {Kull}, {Bachchan},
  {Lacoste-Seris}, {Lanza}, {Lavigne}, {Le Poncin-Lafitte}, {Lebreton},
  {Lebzelter}, {Leccia}, {Leclerc}, {Lecoeur-Taibi}, {Lemaitre}, {Lenhardt},
  {Leroux}, {Liao}, {Licata}, {Lindstr{\o}m}, {Lister}, {Livanou}, {Lobel},
  {L{\"o}ffler}, {L{\'o}pez}, {Lopez-Lozano}, {Lorenz}, {Loureiro},
  {MacDonald}, {Magalh{\~a}es Fernandes}, {Managau}, {Mann}, {Mantelet},
  {Marchal}, {Marchant}, {Marconi}, {Marie}, {Marinoni}, {Marrese},
  {Marschalk{\'o}}, {Marshall}, {Mart{\'\i}n-Fleitas}, {Martino}, {Mary},
  {Matijevi{\v{c}}}, {Mazeh}, {McMillan}, {Messina}, {Mestre}, {Michalik},
  {Millar}, {Miranda}, {Molina}, {Molinaro}, {Molinaro}, {Moln{\'a}r},
  {Moniez}, {Montegriffo}, {Monteiro}, {Mor}, {Mora}, {Morbidelli}, {Morel},
  {Morgenthaler}, {Morley}, {Morris}, {Mulone}, {Muraveva}, {Musella},
  {Narbonne}, {Nelemans}, {Nicastro}, {Noval}, {Ord{\'e}novic},
  {Ordieres-Mer{\'e}}, {Osborne}, {Pagani}, {Pagano}, {Pailler}, {Palacin},
  {Palaversa}, {Parsons}, {Paulsen}, {Pecoraro}, {Pedrosa}, {Pentik{\"a}inen},
  {Pereira}, {Pichon}, {Piersimoni}, {Pineau}, {Plachy}, {Plum}, {Poujoulet},
  {Pr{\v{s}}a}, {Pulone}, {Ragaini}, {Rago}, {Rambaux}, {Ramos-Lerate},
  {Ranalli}, {Rauw}, {Read}, {Regibo}, {Renk}, {Reyl{\'e}}, {Ribeiro},
  {Rimoldini}, {Ripepi}, {Riva}, {Rixon}, {Roelens}, {Romero-G{\'o}mez},
  {Rowell}, {Royer}, {Rudolph}, {Ruiz-Dern}, {Sadowski}, {Sagrist{\`a}
  Sell{\'e}s}, {Sahlmann}, {Salgado}, {Salguero}, {Sarasso}, {Savietto},
  {Schnorhk}, {Schultheis}, {Sciacca}, {Segol}, {Segovia}, {Segransan},
  {Serpell}, {Shih}, {Smareglia}, {Smart}, {Smith}, {Solano}, {Solitro},
  {Sordo}, {Soria Nieto}, {Souchay}, {Spagna}, {Spoto}, {Stampa}, {Steele},
  {Steidelm{\"u}ller}, {Stephenson}, {Stoev}, {Suess}, {S{\"u}veges}, {Surdej},
  {Szabados}, {Szegedi-Elek}, {Tapiador}, {Taris}, {Tauran}, {Taylor},
  {Teixeira}, {Terrett}, {Tingley}, {Trager}, {Turon}, {Ulla}, {Utrilla},
  {Valentini}, {van Elteren}, {Van Hemelryck}, {van Leeuwen}, {Varadi},
  {Vecchiato}, {Veljanoski}, {Via}, {Vicente}, {Vogt}, {Voss}, {Votruba},
  {Voutsinas}, {Walmsley}, {Weiler}, {Weingrill}, {Werner}, {Wevers},
  {Whitehead}, {Wyrzykowski}, {Yoldas}, {{\v{Z}}erjal}, {Zucker}, {Zurbach},
  {Zwitter}, {Alecu}, {Allen}, {Allende Prieto}, {Amorim},
  {Anglada-Escud{\'e}}, {Arsenijevic}, {Azaz}, {Balm}, {Beck}, {Bernstein},
  {Bigot}, {Bijaoui}, {Blasco}, {Bonfigli}, {Bono}, {Boudreault}, {Bressan},
  {Brown}, {Brunet}, {Bunclark}, {Buonanno}, {Butkevich}, {Carret}, {Carrion},
  {Chemin}, {Ch{\'e}reau}, {Corcione}, {Darmigny}, {de Boer}, {de Teodoro}, {de
  Zeeuw}, {Delle Luche}, {Domingues}, {Dubath}, {Fodor}, {Fr{\'e}zouls},
  {Fries}, {Fustes}, {Fyfe}, {Gallardo}, {Gallegos}, {Gardiol}, {Gebran},
  {Gomboc}, {G{\'o}mez}, {Grux}, {Gueguen}, {Heyrovsky}, {Hoar}, {Iannicola},
  {Isasi Parache}, {Janotto}, {Joliet}, {Jonckheere}, {Keil}, {Kim},
  {Klagyivik}, {Klar}, {Knude}, {Kochukhov}, {Kolka}, {Kos}, {Kutka}, {Lainey},
  {LeBouquin}, {Liu}, {Loreggia}, {Makarov}, {Marseille}, {Martayan},
  {Martinez-Rubi}, {Massart}, {Meynadier}, {Mignot}, {Munari}, {Nguyen},
  {Nordlander}, {Ocvirk}, {O'Flaherty}, {Olias Sanz}, {Ortiz}, {Osorio},
  {Oszkiewicz}, {Ouzounis}, {Palmer}, {Park}, {Pasquato}, {Peltzer}, {Peralta},
  {P{\'e}turaud}, {Pieniluoma}, {Pigozzi}, {Poels}, {Prat}, {Prod'homme},
  {Raison}, {Rebordao}, {Risquez}, {Rocca-Volmerange}, {Rosen}, {Ruiz-Fuertes},
  {Russo}, {Sembay}, {Serraller Vizcaino}, {Short}, {Siebert}, {Silva},
  {Sinachopoulos}, {Slezak}, {Soffel}, {Sosnowska}, {Strai{\v{z}}ys}, {ter
  Linden}, {Terrell}, {Theil}, {Tiede}, {Troisi}, {Tsalmantza}, {Tur},
  {Vaccari}, {Vachier}, {Valles}, {Van Hamme}, {Veltz}, {Virtanen}, {Wallut},
  {Wichmann}, {Wilkinson}, {Ziaeepour}, \& {Zschocke}}]{gaia16}
{Gaia Collaboration}, {Prusti}, T., {de Bruijne}, J.~H.~J., {et~al.} 2016,
  \aap, 595, A1, \dodoi{10.1051/0004-6361/201629272}

\bibitem[{{Green} {et~al.}(2019){Green}, {Schlafly}, {Zucker}, {Speagle}, \&
  {Finkbeiner}}]{green19}
{Green}, G.~M., {Schlafly}, E., {Zucker}, C., {Speagle}, J.~S., \&
  {Finkbeiner}, D. 2019, \apj, 887, 93, \dodoi{10.3847/1538-4357/ab5362}

\bibitem[{{Ibata} {et~al.}(2017){Ibata}, {McConnachie}, {Cuillandre}, {Fantin},
  {Haywood}, {Martin}, {Bergeron}, {Beckmann}, {Bernard}, {Bonifacio},
  {Caffau}, {Carlberg}, {C{\^o}t{\'e}}, {Cabanac}, {Chapman}, {Duc}, {Durret},
  {Famaey}, {Fabbro}, {Gwyn}, {Hammer}, {Hill}, {Hudson}, {Lan{\c{c}}on},
  {Lewis}, {Malhan}, {di Matteo}, {McCracken}, {Mei}, {Mellier}, {Navarro},
  {Pires}, {Pritchet}, {Reyl{\'e}}, {Richer}, {Robin}, {S{\'a}nchez-Janssen},
  {Sawicki}, {Scott}, {Scottez}, {Spekkens}, {Starkenburg}, {Thomas}, \&
  {Venn}}]{CFIS_UNIONS}
{Ibata}, R.~A., {McConnachie}, A., {Cuillandre}, J.-C., {et~al.} 2017, \apj,
  848, 128, \dodoi{10.3847/1538-4357/aa855c}

\bibitem[{{Kollmeier} {et~al.}(2017){Kollmeier}, {Zasowski}, {Rix}, {Johns},
  {Anderson}, {Drory}, {Johnson}, {Pogge}, {Bird}, {Blanc}, {Brownstein},
  {Crane}, {De Lee}, {Klaene}, {Kreckel}, {MacDonald}, {Merloni}, {Ness},
  {O'Brien}, {Sanchez-Gallego}, {Sayres}, {Shen}, {Thakar}, {Tkachenko},
  {Aerts}, {Blanton}, {Eisenstein}, {Holtzman}, {Maoz}, {Nandra}, {Rockosi},
  {Weinberg}, {Bovy}, {Casey}, {Chaname}, {Clerc}, {Conroy}, {Eracleous},
  {G{\"a}nsicke}, {Hekker}, {Horne}, {Kauffmann}, {McQuinn}, {Pellegrini},
  {Schinnerer}, {Schlafly}, {Schwope}, {Seibert}, {Teske}, \& {van
  Saders}}]{Kollmeier2017}
{Kollmeier}, J.~A., {Zasowski}, G., {Rix}, H.-W., {et~al.} 2017, arXiv
  e-prints, arXiv:1711.03234, \dodoi{10.48550/arXiv.1711.03234}

\bibitem[{{Lacy} {et~al.}(2020){Lacy}, {Baum}, {Chandler}, {Chatterjee},
  {Clarke}, {Deustua}, {English}, {Farnes}, {Gaensler}, {Gugliucci},
  {Hallinan}, {Kent}, {Kimball}, {Law}, {Lazio}, {Marvil}, {Mao}, {Medlin},
  {Mooley}, {Murphy}, {Myers}, {Osten}, {Richards}, {Rosolowsky}, {Rudnick},
  {Schinzel}, {Sivakoff}, {Sjouwerman}, {Taylor}, {White}, {Wrobel},
  {Andernach}, {Beasley}, {Berger}, {Bhatnager}, {Birkinshaw}, {Bower},
  {Brandt}, {Brown}, {Burke-Spolaor}, {Butler}, {Comerford}, {Demorest}, {Fu},
  {Giacintucci}, {Golap}, {G{\"u}th}, {Hales}, {Hiriart}, {Hodge}, {Horesh},
  {Ivezi{\'c}}, {Jarvis}, {Kamble}, {Kassim}, {Liu}, {Loinard}, {Lyons},
  {Masters}, {Mezcua}, {Moellenbrock}, {Mroczkowski}, {Nyland}, {O'Dea},
  {O'Sullivan}, {Peters}, {Radford}, {Rao}, {Robnett}, {Salcido}, {Shen},
  {Sobotka}, {Witz}, {Vaccari}, {van Weeren}, {Vargas}, {Williams}, \&
  {Yoon}}]{Lacy20}
{Lacy}, M., {Baum}, S.~A., {Chandler}, C.~J., {et~al.} 2020, \pasp, 132,
  035001, \dodoi{10.1088/1538-3873/ab63eb}

\bibitem[{{Norris}(2011)}]{norris11}
{Norris}, R.~P. 2011, Journal of Astrophysics and Astronomy, 32, 599,
  \dodoi{10.1007/s12036-011-9119-z}

\bibitem[{{Slater} {et~al.}(2020){Slater}, {Ivezi{\'c}}, \&
  {Lupton}}]{slater20}
{Slater}, C.~T., {Ivezi{\'c}}, {\v{Z}}., \& {Lupton}, R.~H. 2020, \aj, 159, 65,
  \dodoi{10.3847/1538-3881/ab6166}

\bibitem[{{Taylor} {et~al.}(2018){Taylor}, {Cirasuolo}, {Afonso}, {Carollo},
  {Evans}, {Flores}, {Maiolino}, {Paltani}, {Vanzi}, {Abreu}, {Amans},
  {Atkinson}, {Barrett}, {Beard}, {B{\'e}chet}, {Black}, {Boettger},
  {Brierley}, {Buscher}, {Cabral}, {Cochrane}, {Coelho}, {Colling},
  {Conzelmann}, {Dalessio}, {Dauvin}, {Davidson}, {Drass}, {D{\"u}nner},
  {Fairley}, {Fasola}, {Ferruzzi}, {Fisher}, {Flores}, {Garilli}, {Gaudemard},
  {Gonzalez}, {Guinouard}, {Gutierrez}, {Hammersley}, {Haigron}, {Haniff},
  {Hayati}, {Ives}, {Iwert}, {Laporte}, {Lee}, {Li Causi}, {Luco}, {Macleod},
  {Mainieri}, {Maire}, {Melse}, {Nix}, {Oliva}, {Oliveira}, {Origlia}, {Parry},
  {Pedichini}, {Piazzesi}, {Rees}, {Reix}, {Rodrigues}, {Rojas}, {Rota},
  {Royer}, {Santos}, {Schnell}, {Shen}, {Sordet}, {Strachan}, {Sun}, {Tait},
  {Torres}, {Tozzi}, {Tulloch}, {Navarro}, {Von Dran}, {Waring}, {Watson},
  {Woodward}, \& {Yang}}]{taylor18}
{Taylor}, W., {Cirasuolo}, M., {Afonso}, J., {et~al.} 2018, in Society of
  Photo-Optical Instrumentation Engineers (SPIE) Conference Series, Vol. 10702,
  Ground-based and Airborne Instrumentation for Astronomy VII, ed. C.~J.
  {Evans}, L.~{Simard}, \& H.~{Takami}, 107021G, \dodoi{10.1117/12.2313403}

\bibitem[{{WFIRST Astrometry Working Group} {et~al.}(2019){WFIRST Astrometry
  Working Group}, {Sanderson}, {Bellini}, {Casertano}, {Lu}, {Melchior},
  {Libralato}, {Bennett}, {Shao}, {Rhodes}, {Sohn}, {Malhotra}, {Gaudi},
  {Fall}, {Nelan}, {Guhathakurta}, {Anderson}, \& {Ho}}]{sanderson19}
{WFIRST Astrometry Working Group}, {Sanderson}, R.~E., {Bellini}, A., {et~al.}
  2019, Journal of Astronomical Telescopes, Instruments, and Systems, 5,
  044005, \dodoi{10.1117/1.JATIS.5.4.044005}

\end{thebibliography}
\bibliographystyle{aasjournal}

\end{document}